\documentclass{article}

\usepackage{arxiv}

\usepackage[utf8]{inputenc} 
\usepackage[T1]{fontenc}    
\usepackage{hyperref}       
\usepackage{url}            
\usepackage{booktabs}       
\usepackage{amsfonts}       
\usepackage{nicefrac}       
\usepackage{microtype}      
\usepackage{cleveref}       
\usepackage{lipsum}         
\usepackage{graphicx}
\usepackage{natbib}
\usepackage{doi}

\title{On the Use of Causal Graphical Models for Designing Experiments in the Automotive Domain}

\author{David Issa Mattos \\
	Volvo Cars\\
	Gothenburg, Sweden\\
	\texttt{david.mattos@volvocars.com} \\
	\And
	Yuchu Liu \\
	Volvo Cars\\
	Gothenburg, Sweden\\
	\texttt{yuchu.liy@volvocars.com} \\
}

\renewcommand{\headeright}{Mattos and Liu}
\renewcommand{\undertitle}{}
\renewcommand{\shorttitle}{}

\newif\ifacm
\acmfalse

\hypersetup{
pdftitle={On the Use of Causal Graphical Models for Designing Experiments in the Automotive Domain},
pdfauthor={David Issa Mattos, Yuchu Liu},
pdfkeywords={Experimentation, Causality, Causal Graphical Models, DAG},
}

\begin{document}
\maketitle

\begin{abstract}
	Randomized field experiments are the gold standard for evaluating the impact of software changes on customers. In the online domain, randomization has been the main tool to ensure exchangeability. However, due to the different deployment conditions and the high dependence on the surrounding environment, designing experiments for automotive software needs to consider a higher number of restricted variables to ensure conditional exchangeability. In this paper, we show how at Volvo Cars we utilize causal graphical models to design experiments and explicitly communicate the assumptions of experiments. These graphical models are used to further assess the experiment validity, compute direct and indirect causal effects, and reason on the transportability of the causal conclusions.
\end{abstract}

\keywords{Experimentation, Causality, Causal Graphical Models, DAG}

\section{Introduction}
Randomized field experiments, such as A/B testing, have been extensively used by online companies to assess and validate product change ideas \citep{Fabijan2017}. In the simplest case, users are randomized between two groups: the control (the existing software system), and the treatment (the software system with the desired change) groups. 

The randomization process is a simple and reliable way to allow the control and treatment groups to be exchangeable and to estimate the (unbiased) causal effect of the software change. However, in several practical applications fully randomized experiments are not desirable or even possible to be conducted. In this context, different tools can be used to estimate the causal effect of software changes, such as quasi-experiments, matching, instrumental variables, etc \citep{liu2021size,liu2021bayesian,xu2016evaluating}.

In the automotive domain, several conditions prohibit the use of full randomization experiments in most cases, such as the high degree of interaction of the cyber-physical system with the environment, the deployment limitations, and the limited sample size. Therefore, experiments designed for automotive software development need to be restricted to several confounders that can potentially influence the desired outcome metric.

To address these limitations at Volvo Cars, we utilize causal graphical models \citep{glymour2016causal}, to help design experiments and make the assumptions taken explicit for all. Moreover, these causal models can be used to assess the experiment validity, compute potential direct and indirect effects, and reason about the transportability of the experimental results for other populations. 
\section{Background}

Assessing causality in online experiments is traditionally conducted in the Rubin-Neyman potential outcomes framework \citep{Holland1986}. This framework assesses the causal effect by using counterfactual, what would be the impact of a treatment in a population if it had not been exposed to the treatment. To achieve that, some conditions need to be fulfilled such as positivity (there are samples for both the treatment and the control), exchangeability (there is an independence between the counterfactual outcome and the observed treatment) and consistency (the treatment is the same and well-defined) \citep{hernan2010causal}. While in randomized field experiments positivity and consistency are fulfilled by design, proper randomization is used to achieve exchangeability.

When multiple variables need to be restricted or they cannot be randomized, it is necessary to control for them in the design to ensure conditional exchangeability holds, which means all backdoor paths are blocked in a causal Directed Acyclic Graph (DAG) \citep{hernan2010causal, glymour2016causal}. For this reason, we utilize graphical causal models based on DAG to aid the design of experiments in automotive software engineering. Several books are dedicated to review of graphical causal models, its relation to the potential outcomes framework, and its applications in different areas of science \citep{hernan2010causal, glymour2016causal}.
\section{Using causal graphical models}
Experiments conducted in automotive software development have many conditions that need to be restricted or cannot be randomized. While we have explored some of them in our previous work \citep{liu2021bayesian,liu2021size,xu2016evaluating}, in this paper, we utilize graphical causal models to improve the design and communication of experiments. 

Our general process for using graphical causal models consists of:

\begin{enumerate}
    \item Value mapping: identification of the different aspects (which we call values) the change is expected to impact, such as overall evaluation criteria, guardrails, confounders etc.
    \item Causal mapping: utilizing the subsystems domain knowledge, we create a graphical causal model that maps how the change impacts the different systems and subsystems related to the mapped values. In this step, we differentiate which variables are the main metrics, guardrails, intervention, latent/ non-observable, and other measurable metrics.
    \item Causal mapping validation: The causal map is validated in an iterative  two-stage process. We first review the causal map with a group of experts followed by a consistency check with existing collected data (e.g., checking if the conditional independence that the graph includes holds). These two steps are iterated until we reach a consensus on the quality of the causal map. In other words,  we combine the knowledge-driven and data-driven causal discovery processes in our practice.
    \item Experimental design and validity: based on the causal map we can start restricting variables that cannot be randomized. Utilizing algorithms for identifying conditional exchangeability in a DAG. \citep{glymour2016causal,hernan2010causal}, we can verify which variables are required to be controlled in the experiment design. When designing the experiment and determining how the treatment assignment process will occur, such as a combination of randomization and restricted variables, additional conditional independence relations will arise that help verify the validity of the design. These conditional independence relations are an extra check similar to the results of an A/A test and different sample-ratio-mismatch criteria (which are also derived automatically from the DAG).
    \item Analysis: after the experiment, data is collected. We query the DAG to guide the analysis, for instance, we might be interested in separating the direct and the indirect effects, as opposed to the total effect obtained in the experiment, as well as evaluating causal transportability questions \citep{hernan2010causal}.
\end{enumerate}

We provide below an illustrative example of simplified case conducted at Volvo Cars in Figure \ref{fig:dag}. In this example, a new software modification on the climate system was aimed at reducing energy consumption (the direct effect). However, the new software could potentially affect how users interact with the climate system and generate a potential indirect effect of increasing energy consumption. The causal diagram also contains latent and non-measurable variables.

\ifacm
    \begin{figure}[t]
    \centering
    \includegraphics[width=3.0in]{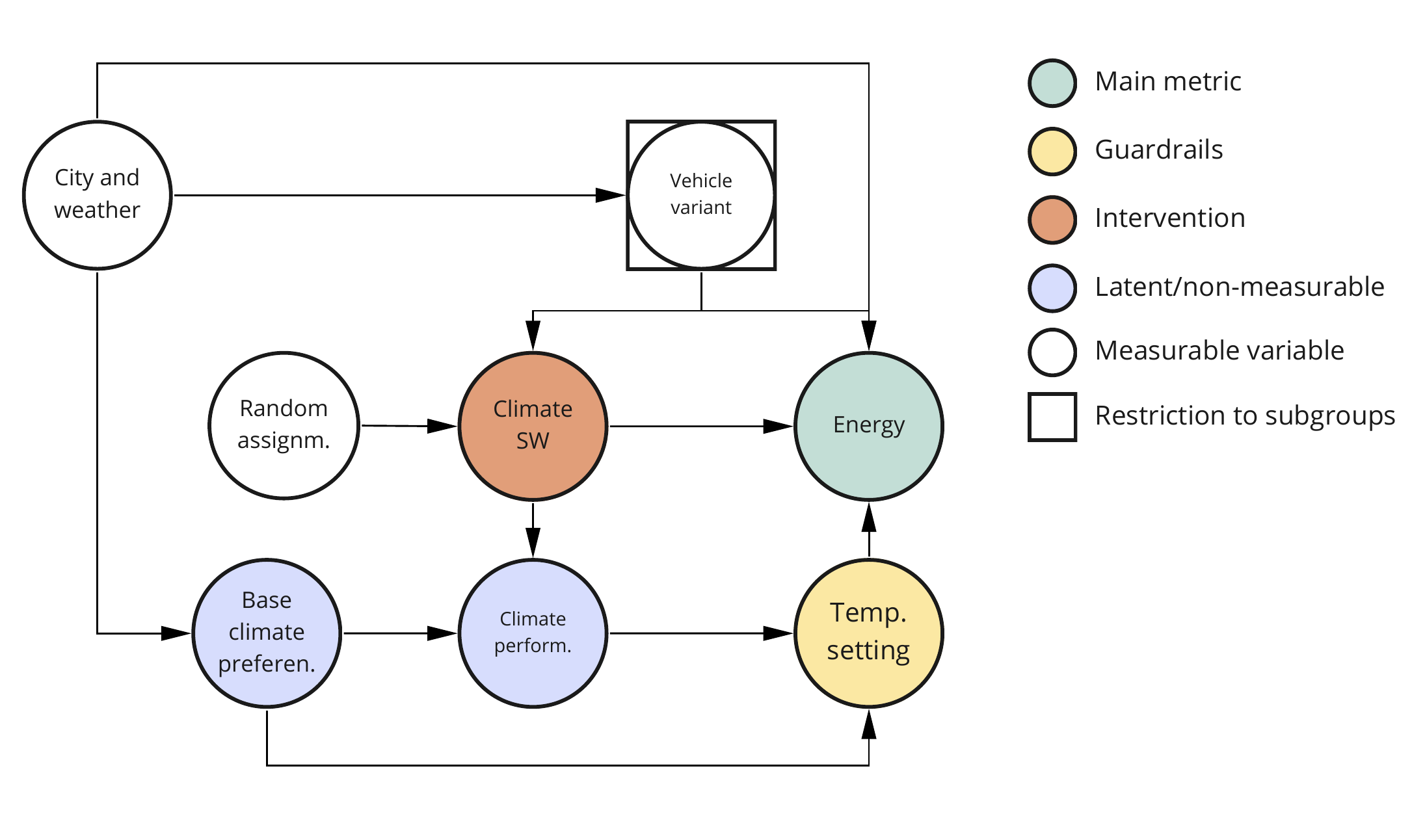}
    \caption{A simplified causal graphical model.}
    \label{fig:dag}
\end{figure}
\else
    \begin{figure}[t]
    \centering
    \includegraphics[width=\textwidth]{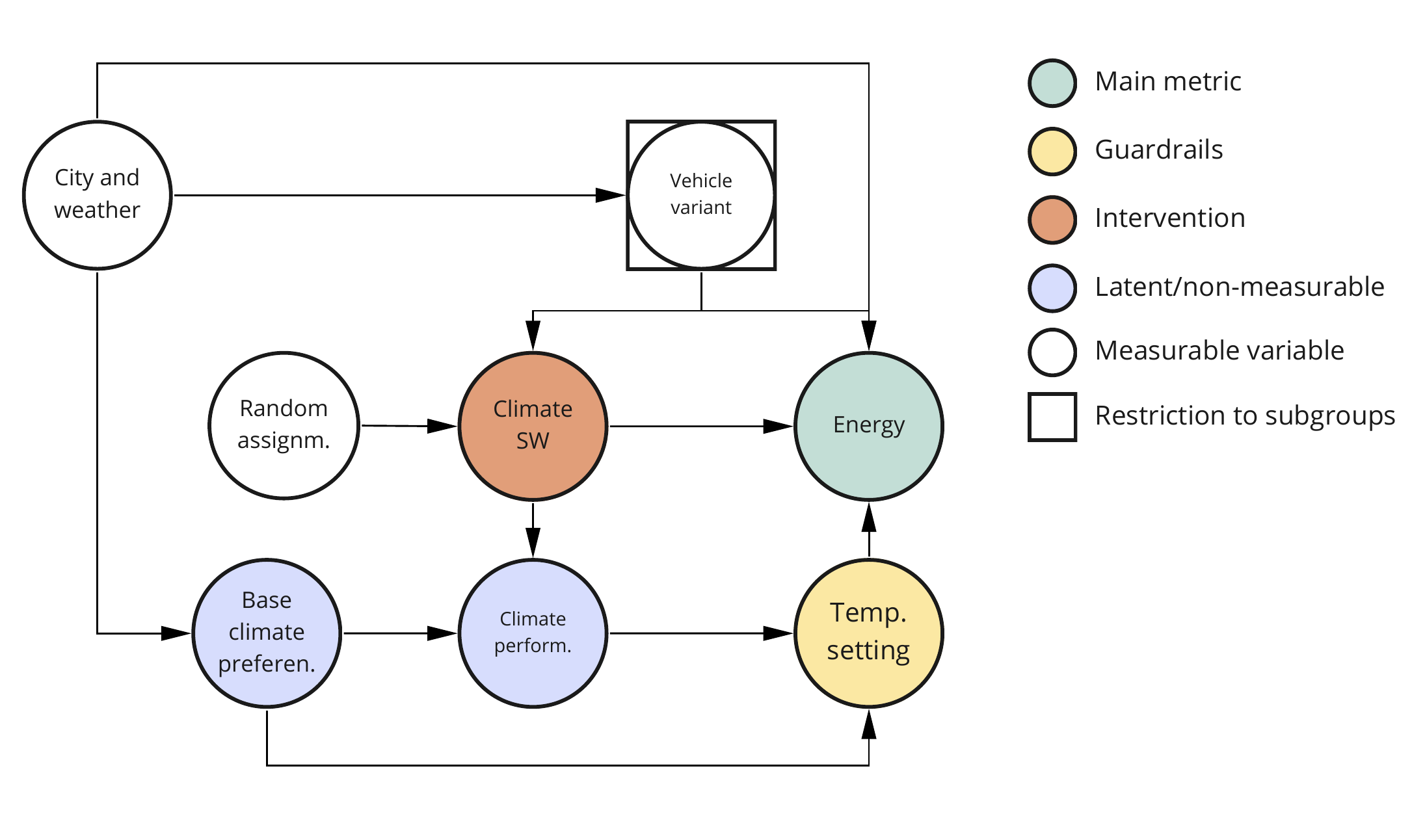}
    \caption{A simplified causal graphical model.}
    \label{fig:dag}
\end{figure}
\fi

Using this causal graph, we could find the necessary adjustment sets of confounding factors (controlling for the vehicle variant) required to identify the unbiased total causal effect of the software change that is the result of the A/B test. For example, if we want to identify the direct effect, we need to adjust for the city, temperature, and vehicle variant. Assuming linearity, the indirect effect, the potential degrading effect of the climate software, can be calculated by subtracting the direct effect from the total effect. 

We can control for the adjustment sets by identifying the conditional causal effect by strata, adjusting it through inverse probability weighting, or if assuming linearity of the causal effects we can add the variable in the linear model. The following link contains a short appendix on how the analysis of this example was conducted: \url{https://davidissamattos.github.io/ease-2022-causal/}.

The graphical model from Figure \ref{fig:dag} is used internally to communicate the experiment assumptions between different teams which increases transparency of the experiment design and analysis decisions. The use of graphical models serves as a boundary object to facilitate the communication interface between several teams \citep{heynstructural, wohlrab2019boundary}.
\section{Conclusion}
Causal models are a powerful tool to assess causality in any application. It is general enough to encompass and leverage experiments, quasi-experiments, and observational studies in a single consistent framework. The main disadvantage of such a framework is the need to construct a correct causal graphical model and the real causal structure might be hard or impossible to obtain in certain cases. However, by combining tools for automatic causal discovery from data with domain knowledge, we believe we can provide a meaningful and actionable causal graphical model for most applications.


\bibliographystyle{unsrtnat}
\bibliography{bibliography/refs.bib} 
\end{document}